\documentclass[12pt]{article}
\usepackage{epsfig}
 
\begin{document}
\thispagestyle{empty} 
 
\begin{flushright}
{\small
SLAC--PUB--8010--REV\\
January 1999\\}
\end{flushright}
\bigskip
 
\Large
\bf
\begin{center}
Certainty and Uncertainty in the Practice of Science: Electrons, Muons, and Taus
\end{center}
\medskip
\normalfont
\normalsize
\begin{center}
M.~L. Perl\\
Stanford Linear Accelerator Center, Stanford University,
Stanford, CA 94309\\
\end{center}
 
\bigskip
 
\begin{abstract}
\noindent
\vspace{-4ex}
\begin{quote}
During the past one hundred  years three related elementary particles--the
electron, the muon, and the tau--were discovered by very different scientific
techniques.  The author, who received the Wolf Prize and the Nobel Prize for
the discovery of the tau, uses this history to discuss certainty and uncertainty
in the practice of science.  While the emphasis is on the practice of scientific
research, the paper also explains for the non-physicist some basic ideas in
elementary particle science.
\end{quote}
\end{abstract}
 
\vfill
 
\begin{center}
To be published in {\it Science at the Turn of the Millennium}
\end{center}
\bigskip
\bigskip 
 
\begin{center}
\renewcommand{\baselinestretch}{.5}
\footnotesize
Stanford Linear Accelerator Center,
Stanford University, Stanford, CA 94309
\rule[.5ex]{\textwidth}{.2mm}
Work supported by Department of Energy contract DE--AC03--76SF00515.
\end{center}
\renewcommand{\baselinestretch}{1}
\normalsize

\subsection*{Certainty and uncertainty in the practice of science.}

While science is several thousand years old, it is in the last hundred
years that the practice of science has become tremendously important
in our lives: in the economy, in the technology of war, in the state of
the natural environment, in the condition of our health and in all the
material aspects of our lives.  Many of our thoughts about the next
millennium, our hopes and our fears, have to do with what the findings of science will do for us and what the findings of science will do to us.  We try to predict what these findings of science might be; we want to reassure ourselves that we can control science and that we can direct the practice of science to desirable goals.  There are many goals: some hope for major improvements in material comforts, others hope for the salvation of the natural environment, still others hope for lives without illness and with increased longevity.  These hopes are based on assumptions that the directions of science can be controlled or planned, that there is coherence in the practice of science, that scientists know where their research is going, that any puzzle or problem in the natural world can be solved by enough scientific effort.

I have been a working scientist, an experimenter in physics, for almost fifty years \cite{perlref}, and I am uncomfortable with these assumptions because the practice of science is an uncertain human activity.  Is this a fruitful research direction?  Can this problem be solved?  Are we smart enough or lucky enough to solve the problem?  Do we have the required research technology and if not, can we develop it?  What are our motivations for doing this research?  Will the results of this research have applications?  Will these applications be beneficial or harmful?  

It is best to replace these abstractions by giving the history of one field of science.  I choose the field I know best, the science of elementary particles and in particular, the science of the lepton family of particles.  As I will explain, leptons are, or at least seem to be, very simple elementary particles; thus research on leptons is easy to describe and to use as an example.  

The history of lepton physics is also an apt example because this physics is about 100 years old.  In the middle 1890's Thomson elucidated the nature of the electron \cite{pais,dahl}, the first identified elementary particle and the first lepton.  Since then two heavier electron-like particles were discovered, the muon about mid-century, and the tau, discovered by my colleagues and myself about twenty years ago \cite{perlprix}.  Thus the twentieth century is spanned by the scientific work on the electron, muon, and tau, plus the work on closely associated elementary particles called neutrinos.

As a former United States president was fond of saying, I want to make one thing perfectly clear.  The uncertainties in the practice of science do not necessarily lead to uncertainties in the findings of science.  If experimental results or observations on a phenomenon are verified by other experimenters, if there is logical understanding of the results or observations, then in my philosophy we have learned something real about the natural world.  I am an engineer turned physicist and I have no interest in those philosophies of science that are concerned \pagebreak with whether we do or can know reality.  Similarly I do not believe that the uncertainties in the practice of science will lead to the ``end of science" \cite{horgan}.  I am not of that school.

\subsection*{A note on elementary particles for non-physicists.}

\begin{figure}[t!]
\centerline{
\epsfig{figure=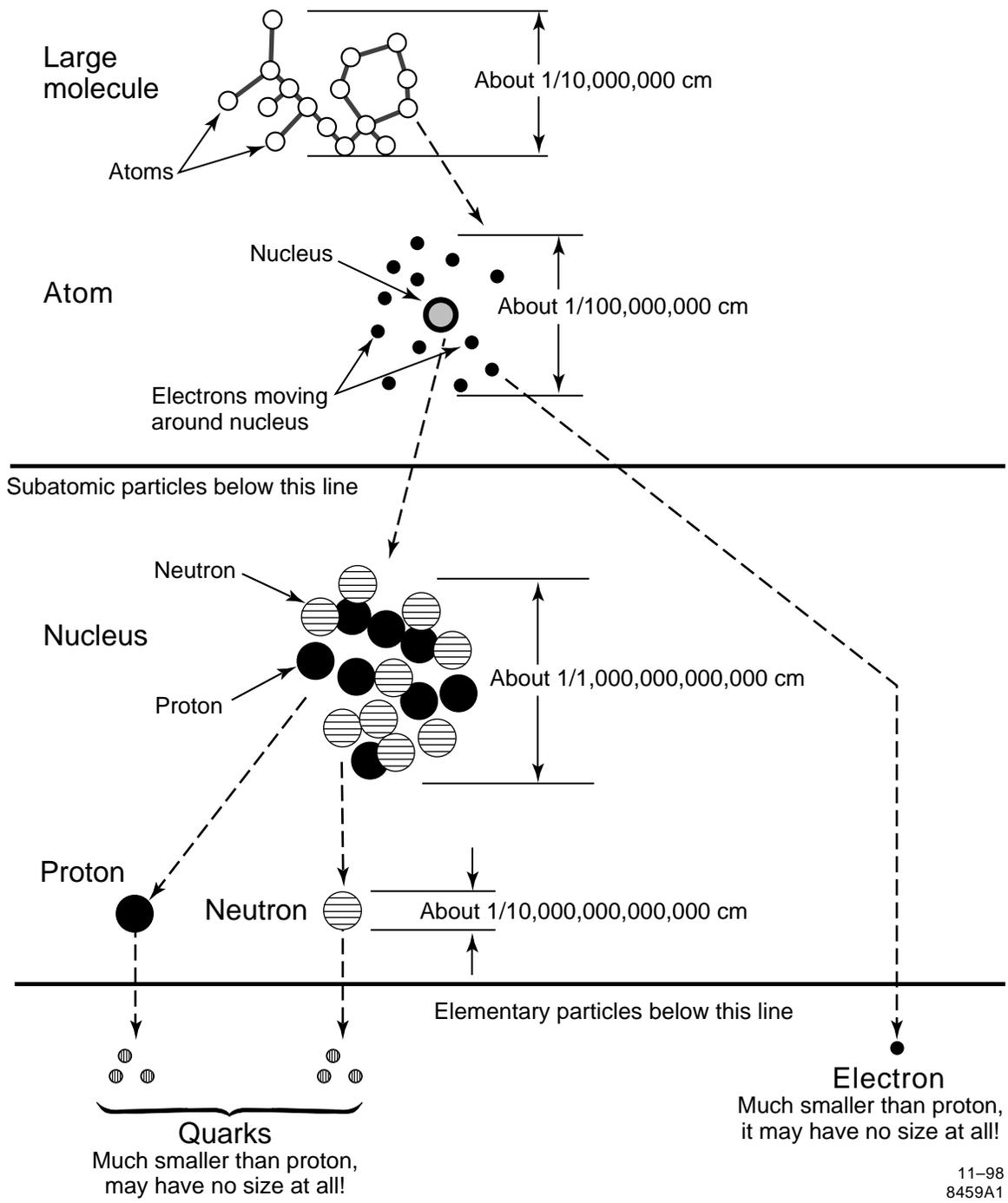}}
\caption{Hierarchy of matter.  At the top are the molecules and
atoms that compose the materials of everyday life.  At the bottom
are the smallest particles of matter that we have so far  found,
the quark and electron elementary particles.}
\label{fig:hierarchy}
\end{figure}

The following are some paragraphs about elementary particles \cite{kane}.  Figure~\ref{fig:hierarchy} shows the hierarchy of matter with the largest kinds of matter, the molecules, at the top.  At the bottom of Figure~\ref{fig:hierarchy} are the elementary particles, the smallest pieces of matter that we have been able to find, smaller than an atom, smaller than a nucleus, less than $10^{-17}$ centimeters in extent; perhaps having no detectable size.  The number $10^{-17}$ means 1/100,000,000,000,000,000 with 17 zeros in the denominator.  This notation for large numbers is a great convenience and I explain it in the Appendix\ref{appendix}.  

Returning to Figure~\ref{fig:hierarchy}, the materials of everyday life such as water and wood and plastics and plant tissue are composed of molecules; and as you know from chemistry and biology, molecules are composed of atoms.  Other materials such as iron and silicon are directly composed of atoms.  But atoms are not simple entities, they themselves are complex, consisting of electrons moving around a nucleus.  

Continuing to move downward in Figure~\ref{fig:hierarchy}, the electron, as far as we know, is not composed of anything else; we cannot break up the electron or find anything inside of it.  The electron is the most prevalent example of an elementary particle.  

On the other hand a nucleus is not simple and is not elementary; a nucleus is made up of protons and neutrons.  At one time neutrons and protons were thought to be elementary particles, but we now know that they are made up of quarks.  As far as we know, quarks like electrons are not composed of anything else; we cannot break up quarks or find anything inside of them.  Thus we have arrived at the bottom of Figure~\ref{fig:hierarchy} and to the simplest particles that compose everyday matter.

Of course these elementary particles, quarks and electrons, may not be so simple; with new ideas and new experimental technology, we may find a deeper structure in these particles.  {\it In the practice of science present understanding may be replaced by a deeper future understanding; but until that replacement occurs we require that present understanding fit existing data.}  A popular and well-advertised speculative theory holds that elementary particles are manifestations of different vibrations of extremely small strings \cite{davies}.  But there is no experimental proof of the validity of the string theory hypothesis.

\begin{figure}[t!]
\centerline{
\epsfig{figure=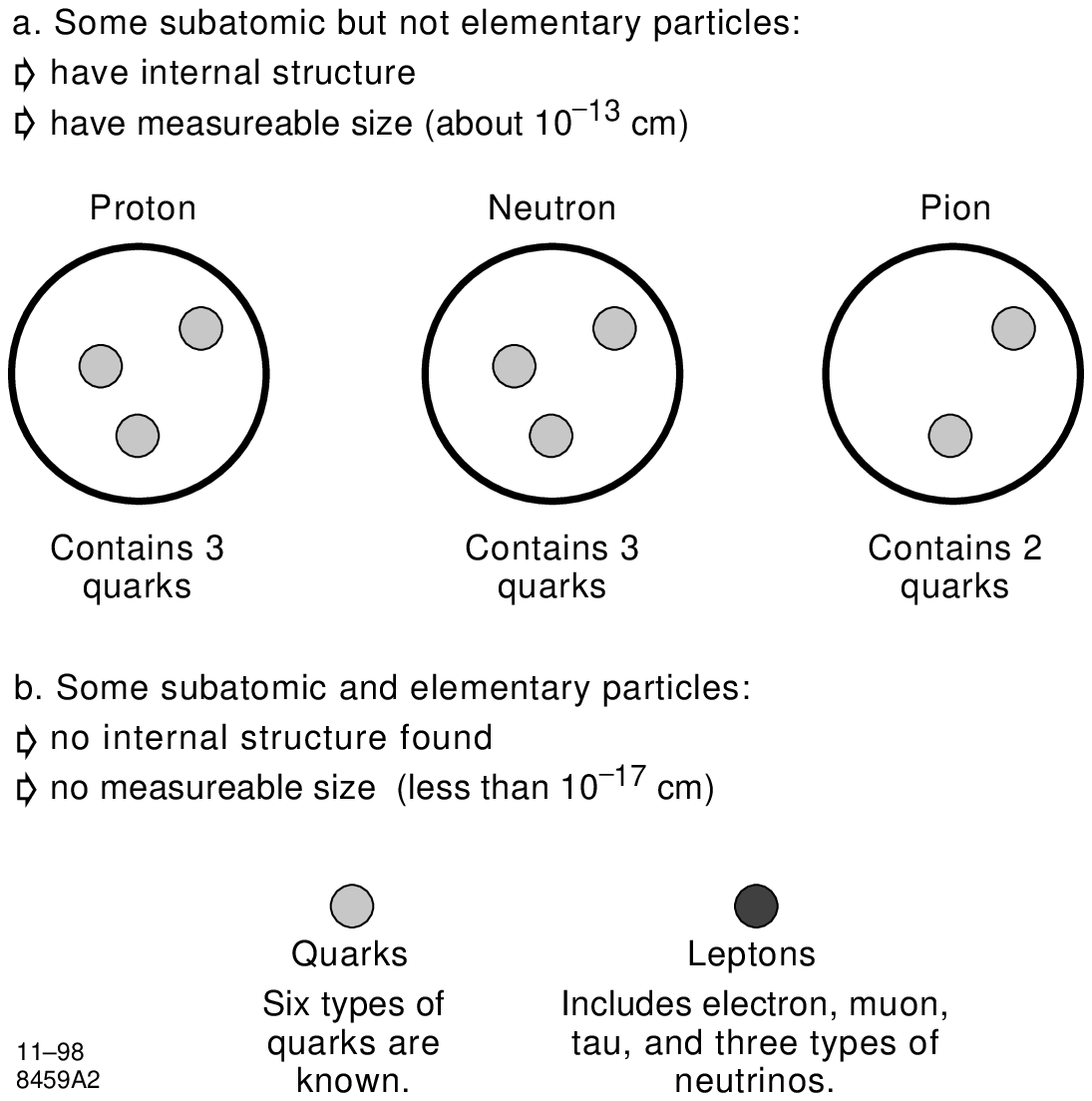}}
\caption{
Two types of subatomic particles: (a) those that are not elementary such as
the proton, neutron, and pion; and (b) those that are elementary such as the
quarks and leptons.
}
\label{fig:twotypes}
\end{figure}

A bit of terminology.  Every particle inside the atom or smaller than the atom is called subatomic.  Nuclei, the neutrons and protons that make up the nucleus, the quarks that make up the neutrons and protons, and the electron are all subatomic particles.  The name elementary is reserved for those subatomic particles that we think are the simplest, those that we think are not made of anything else.  Figure~\ref{fig:twotypes} is my attempt to sort out these distinctions for the reader.  

Electrons and quarks are the elementary particles that exist in everyday matter, but they are not the only elementary particles.  Other elementary particles exist in nature, for example muons and neutrinos exist in the atmosphere and in outer space.  Other elementary particles such as the tau and other quarks can be artificially created.  But this is getting ahead of this history.  

Elementary particles are not just isolated pieces of matter that have nothing to do with each other.  They pull and push on each other and interact with each other, sometimes changing into other kinds of particles.  These interactions occur through four different forces: electromagnetic, weak, gravitational, and strong.  Only two of these forces are of immediate concern.  The electromagnetic force is just the electric and magnetic force that is manifest around us; it is the force involved in electric motors, in electronics, in the behavior of static electricity, in the behavior of lightning.  If an elementary particle has electric charge it is acted upon by the electromagnetic force.  

The strong force is the force that holds the quarks inside the protons and neutrons, and it also holds the nucleus itself together.  The strong force is the basis for the production of energy in our sun, in the stars, and in nuclear reactors.  Unfortunately it is also the basis for the devastating release of energy and radioactivity by atom and hydrogen bombs.

The elementary particles are classified into three families. Two of these families, the leptons and the quarks, are delineated in Table~\ref{table:leptons}.  Leptons do not interact through the strong force, and this decisively separates them from the quarks.  The strong force between quarks compels them to be buried in complicated particles such as protons and neutrons and pions, Figure~\ref{fig:twotypes}.  We have never succeeded in making or finding a single quark isolated by itself.  It is difficult to study the properties of quarks and even more difficult to explain their properties and behavior in simple terms.  
\begin{table}[t]
\begin{center}
\caption{
\label{table:leptons}
Definition of leptons and the differences between leptons and quarks.
For a fuller discussion of leptons and quarks see \protect\cite{kane}.
} 
\vspace{1ex}
\begin{tabular}{|p{.3\textwidth}|p{.3\textwidth}|p{.3\textwidth}|}
\hline
Property&Lepton&Quark\\
\hline
\hline
Acted upon by the strong force?&No&Yes\\
\hline
Can be isolated as a single particle?&Yes&Never observed, therefore taken as no\\
\hline
Acted upon by the electromagnetic force?&Yes, if charged&Yes\\
\hline
Electric charge in units of $1.6 \times 10^{-19}$ coulombs&$+1,-1,$ or 0&$+2/3,-2/3,+1/3,$\linebreak or $-1/3$\\
\hline
\end{tabular}
\end{center}
\end{table}

Conversely leptons, free of the strong force, can be isolated and studied individually.  It is also easy to explain their properties in simple terms.  This is why I have devoted much of my research to leptons and why the history of their discovery has pleasing simplicity.  

There is a third class of elementary particles that will not concern us: the particles that carry the basic forces. (The idea of a force being carried by a particle is a quantum mechanical concept.)  For the sake of completeness these particles are the gluon that carries the strong force; the photon that carries the electromagnetic force; and the $W$ and $Z$ particles that carry the weak force, a force I have not discussed.  If quantum mechanics can be applied to the gravitational force in the same way that it is applied to the other forces, then there is another particle called the graviton that carries the gravitational force.  

I will keep my particle physics discussions simple, and to do this I will ignore distinctions that are irrelevant to the matter at hand.  For example there is no need in this paper to distinguish between particles and antiparticles \cite{kane}, and so neutrinos and antineutrinos are simply called neutrinos, quarks and antiquarks are simply called quarks.

\subsection*{Classic science: cathode rays and the discovery of the electron.}

The discovery of the electron is a classic example of scientific discovery \cite{pais,dahl}. Classic in how the effort to understand the phenomenon called cathode rays led to the electron's discovery; classic in how so much was explained once the electron's properties were measured; and classic in how the applications of basic research on the electron has led to radio, television, transistors, computers, and who knows what next.

\begin{figure}[t!]
\centerline{
\epsfig{figure=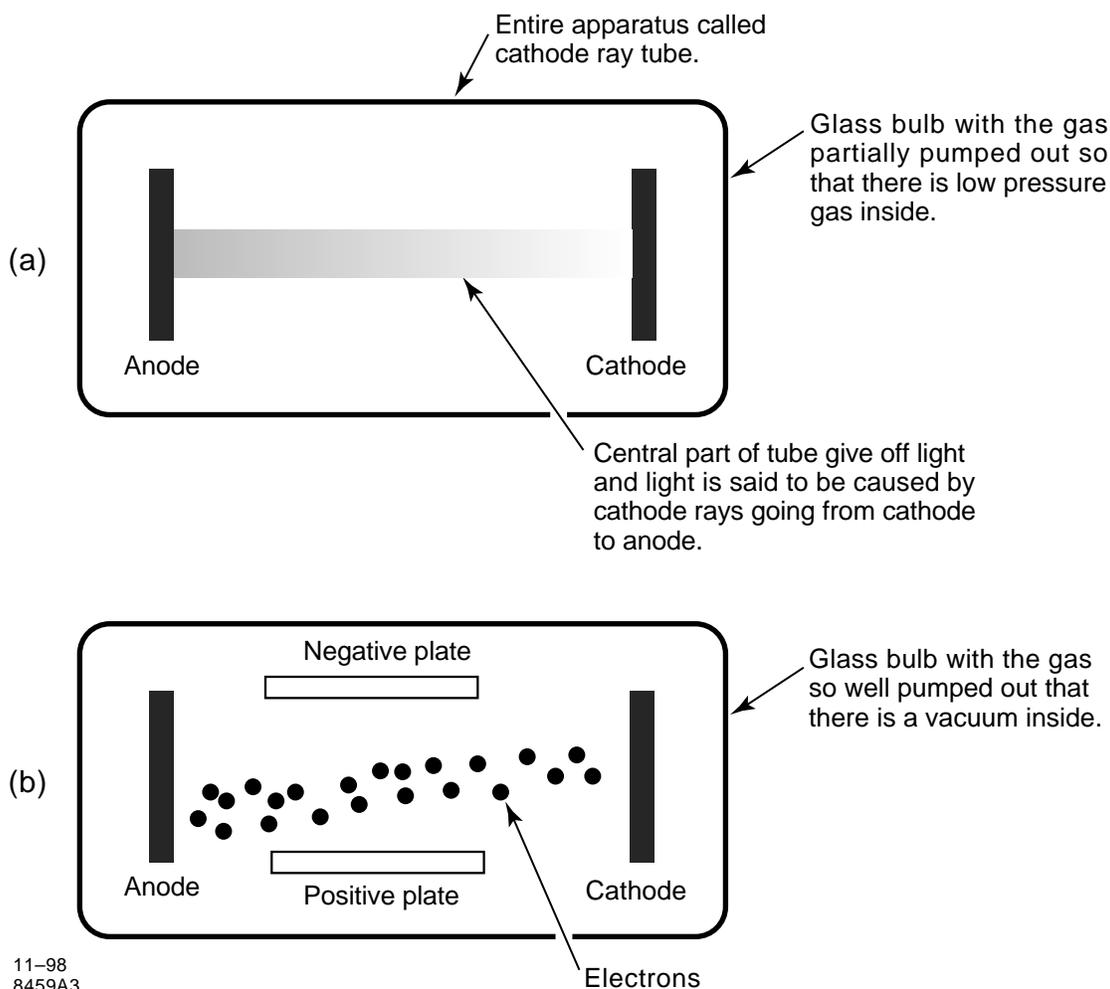}}
\caption{
The electron nature of cathode rays.  (a) A cathode ray tube with low gas
pressure in which the rays, now known to be electrons, cause the gas to emit light.
  (b) The Thomson experiment demonstrating that in a tube with a good vacuum the
electrons, as they move from the cathode to the anode, are deflected by a
perpendicular electric field.
}
\label{fig:cathode}
\end{figure}

It was already known in the eighteenth century that an electrical voltage applied between metal plates in a partially evacuated glass tube could produce light.  Inside the tube the gas glowed; the size, shape, and color of the glowing region depended on the voltage, gas pressure, and shape of the tube.  This phenomenon was called a cathode ray because the light seemed to be caused by rays coming from one of the metal plates inside the tube, specifically from the plate having negative charge, the cathode, Figure~\ref{fig:cathode}.  We see the same phenomenon today in neon lights.  Television picture tubes and computer monitors are also cathode ray tubes, although in these devices the gas pressure is very small.  

Many physicists of the late nineteenth century studied the cathode ray phenomenon, including famous names such as Crookes, Hertz, and Thomson.  Gradually more and more was learned experimentally about cathode rays.  For example it was learned that the rays are bent by a magnetic field and that the rays either carry, or cause the transfer of, negative electric charge.  Still until the middle 1890's there was dispute about the nature of cathode rays.  Some physicists took the rays to be made up of negatively charged matter, the particles we now call electrons.  Others believed the rays to be a kind of electromagnetic wave.  There were several objections to the particle explanation.  The most substantial objection was that the rays should bend in an electric field if they are charged particles, but this bending had not been observed.

The dilemma was resolved in 1895 by Thomson using an improved vacuum pump.  Thomson demonstrated that in a cathode ray tube with a sufficiently good vacuum, the cathode rays were bent in an electric field \cite{pais,dahl}.  A good vacuum is one in which just about all the gas in the tube has been removed.  Describing his experiment with the tube shown in Figure~\ref{fig:cathode}b he wrote, ``At high exhaustion the rays were deflected when the two aluminum plates were connected with the terminals of a battery of small storage cells...  The deflection was proportional to the difference of potential between the plates....  It was only when the vacuum was a good one that the deflection took place.''

Earlier attempts to deflect cathode rays in an electric field had failed because there was still gas in the tube and there was electrical conduction in the partial vacuum.  Gas ions collected on the electrical plates, canceling the charges on the plates and therefore canceling the electric field.  Thus the discovery of the electron depended on the gradual improvement of late nineteenth century instrument technology, particularly vacuum pump technology.  {\it Advances in scientific knowledge often depend upon improving the technology used in the practice of science.}

So we see a triumphant discovery after decades of research on cathode rays.  But we also see that this was not a straight march to success.  About half of the experimenters held the wrong idea about the nature of cathode rays for several decades.  {\it This is an important lesson about the practice of science: wrong ideas may persist for a long time.} Today, one hundred years later, we have much better experimental equipment, but we are no smarter.  Today there are similar controversies about observed phenomena ranging from cosmology to biology.  Some of these controversies may be settled soon by discoveries as clear as the discovery of the electron, some may not be settled for a long time. {\it A major uncertainty in the practice of science is when a particular controversy will be settled.} Thomson received the Nobel Prize for settling the cathode ray controversy.

\subsection*{What we know about the electron.}

The process of discovering the electron was interwoven with the process of determining the basic properties of the electron.  By 1911, Millikan \cite{pais} had measured the size of the electric charge of the electron and had shown, within his experimental errors, that all electrons have the same electric charge.  And by the middle 1920's it was known that the electron acts as though it is a perpetually rotating top and as though it is a very small bar magnet.  I have written ``acts as though'' because if the electron has no size, one cannot picture what is rotating or how it can be a magnet.  

The values of the mass and the charge of the electron illustrate how small elementary particles are compared to the objects used in daily life.  The mass of the electron is about $10^{-27}$ grams.  By the way, mass is called weight in everyday language.  A standard size aspirin has a mass of about 1/3 of a gram.  Thus it would take $10^{27}$ electrons to have about the same weight as three aspirins.

The charge of the electron is $1.6 \times 10^{-19}$ coulombs.  In everyday life we don't use the coulomb unit of charge.  We use a unit, however, for the electric current through a wire, the ampere, and electric current is simply the flow of electrons through a wire.  A 100~watt light bulb uses about one ampere of current.  To the nearest factor of ten, one ampere means $10^{19}$ electrons are flowing through the wire per second.  Thus like the electron mass, the electron charge is very small compared to the electrical quantities that occur in everyday life.

\subsection*{Limited knowledge: what we don't know about the electron.}

A physicist living in the early twentieth century and doing research on the electron would probably have believed that we would continue to learn more and more about the electron as the century progressed.  We have indeed learned more and more about how the electron {\it behaves} in metals, semiconductors, and molecules.  We have indeed measured the {\it known properties} of the electron with more and more precision: its mass, charge, and magnetic properties.  But we have made no progress in understanding what sets the mass of the electron.  We have made no progress in understanding why all the known elementary particles with electric charge have charges that are either equal to plus or minus the charge on the electron or are equal to 1/3 or 2/3 of that charge.  All we know is that no elementary particles with other electric charges have been found.  

Thus a research direction that must have seemed obvious and fruitful in the 1920's, research to further uncover the inner nature of the electron, has not progressed.  We keep trying to break up the electron to find its inner nature and we keep trying to find an unexpected property of the electron.  No one knew what further to do in the 1920's and no one knows what else to do now. {\it Uncertainty about the future of a direction in research is a major uncertainty in the practice of science.  Will the direction pay off or will it be fruitless?}

\subsection*{A note about protons, neutrons, and decaying particles.}

The second subatomic particle to be found was the proton. Its discovery and the first measurements of its properties occupied about 1900 to 1920.  We now know that it is not an elementary particle; as shown in Figure~\ref{fig:twotypes} it is made up of three quarks.  Thus the proton differs from the electron in that the proton has an internal structure, while the electron, to the best of our knowledge, has no internal structure. There are two other major differences between the proton and the electron.  First the proton is almost 2000 times heavier.  Second the proton, having a diameter of about $10^{-13}$~cm, is much larger than the electron.  On the other hand there is some similarity: the proton has the same size electric charge as the electron, but the proton is positively charged while the electron is negatively charged.  

Thus by the end of the first quarter of the twentieth century, two apparently fundamental particles of matter were known, the proton and the electron.  And from quantum mechanics it was also known that light could also be considered to be made of particles, called photons.  Thus nature seemed to be presenting us with a beautifully simple system of three particles composing everything.  Unfortunately the world, even on this simplest level, is a lot more complicated.  {\it In the practice of science we sometimes mistake simplicity for truth; nature may be simple or may be complex.}

\begin{figure}[t!]
\centerline{
\epsfig{figure=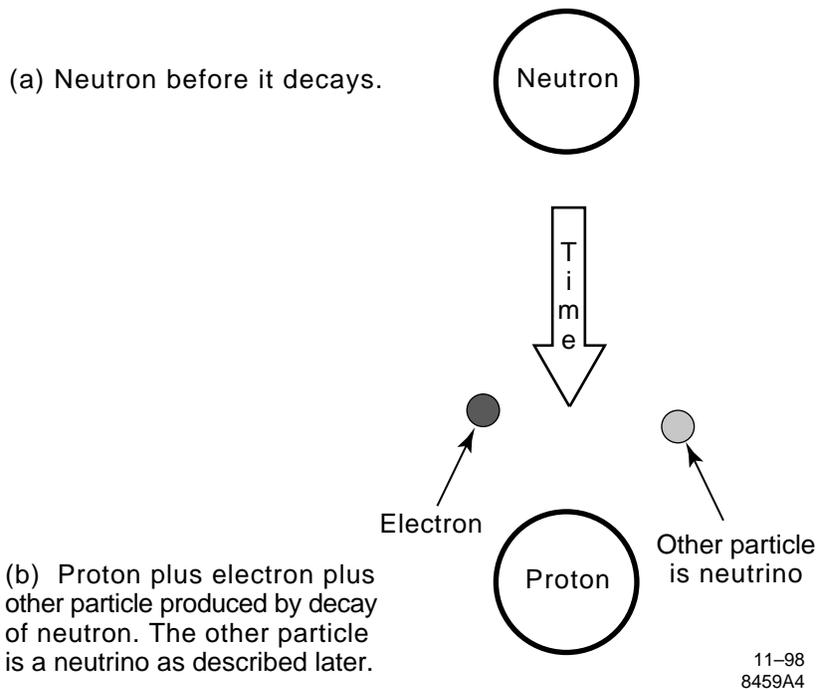}}
\caption{
The decay of a neutron. (a) The neutron before it decays, (b) the particles
produced by the decay.
}
\label{fig:neutron}
\end{figure}

In the early 1930's another subatomic particle, the neutron, was discovered.  The neutron, like the proton, is made out of quarks (Fig.~\ref{fig:twotypes}), but it has zero electric charge.  The neutron is slightly heavier than the proton by about 1/10 of a percent--small difference, but enough to cause a decay process that is common among subatomic particles.  A neutron left to itself does not last forever.  In an average time of about 15 minutes, a neutron spontaneously breaks up into a proton plus an electron plus another elementary particle, the extra mass of the neutron being used to produce the other particles, Figure~\ref{fig:neutron}.  A shorthand to describe the decay process is
$$
\rm
neutron \rightarrow proton + electron + another \  particle.
$$
This means that the particle on left side of the arrow disappears, changing to the particles on the right side of the arrow.  Incidentally as far as we know protons and electrons never decay; left alone, they last forever.

\subsection*{The uncertain road to scientific certainty: cosmic rays and the discovery of the muon.}

Now it is time for me to return to my main story and describe the discovery of the next elementary particle, the muon.  The discovery story begins in the early 1900's with investigations of a natural phenomenon, {\it cosmic rays}, which are not related to cathode rays.  The only connection is linguistic: a ray means something or a group of things moving through space or material in a more or less straight line.  As with the electron, the muon discovery process was interwoven with the process of determining properties; and as with the electron many physicists were involved in these processes. 

\begin{figure}[t!]
\centerline{
\epsfig{figure=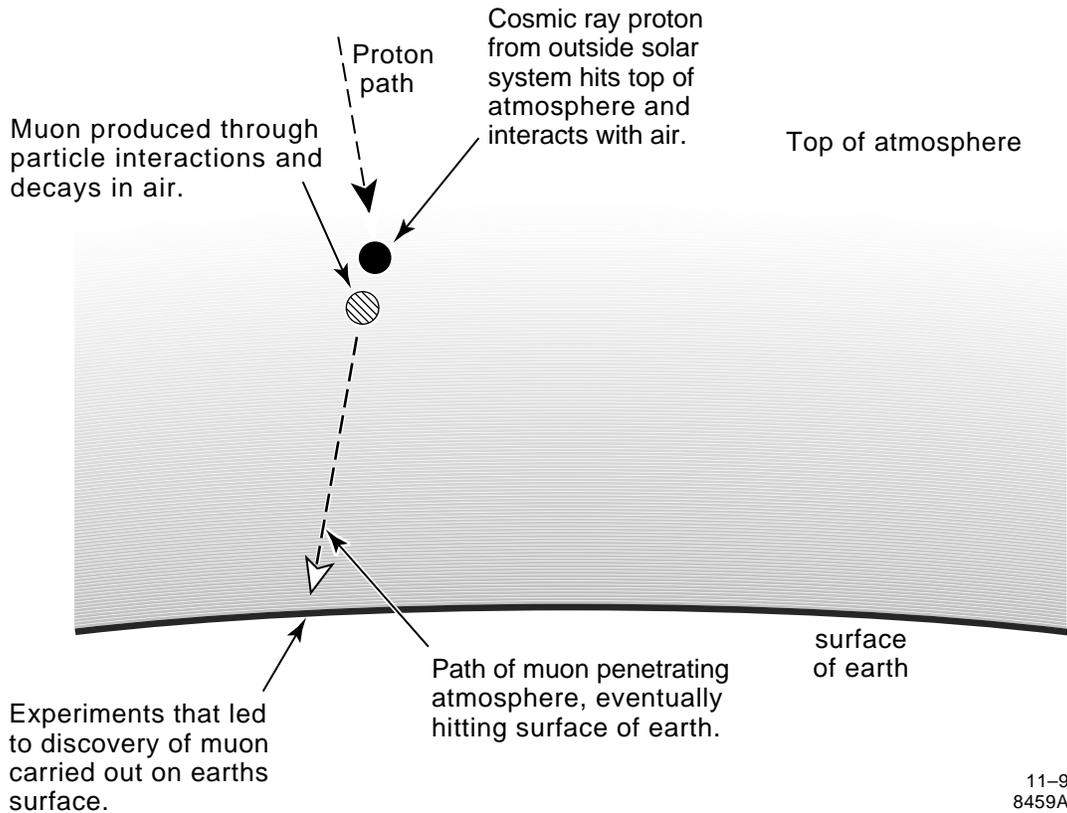}}
\caption{
Cosmic rays and the discovery of the muon.
}
\label{fig:muondis}
\end{figure}

\begin{figure}[p!]
\centerline{
\epsfig{figure=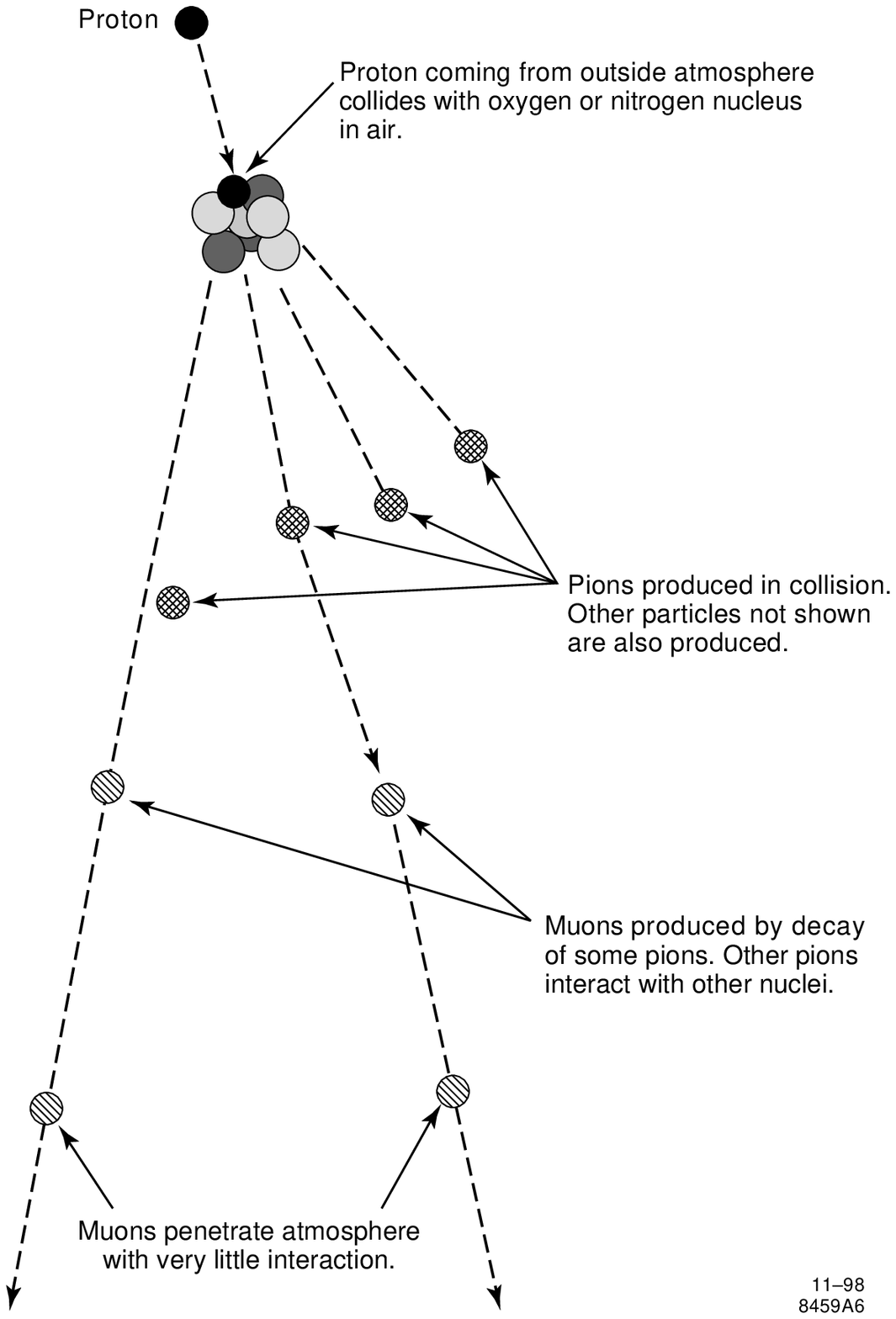}}
\caption{
The process by which muons are produced in cosmic rays.
}
\label{fig:muoncos}
\end{figure}

As we now know, but as was not known in the 1920's, cosmic rays are subatomic particles that enter the Earth's atmosphere traveling with high energy, Figure~\ref{fig:muondis}.  Some are protons and some are atomic nuclei.  Cosmic rays come from outside the solar system and some may come from outside our galaxy.  As cosmic rays pass through our atmosphere, they collide with the oxygen and nitrogen molecules in the air, breaking up the molecules and interacting with the oxygen and nitrogen nuclei to form other particles, mostly pions, Figures~\ref{fig:twotypes} and \ref{fig:muoncos}.  

Returning to the 1910's and 1920's, before all this was known, the first observed effect of cosmic rays was the discovery that the atmosphere could slightly conduct electricity.  Observations also showed that the conductivity extended through the entire depth of the atmosphere, not just at the top of atmosphere.  It was known from research on electrical conductivity in gases, research by the way closely tied to cathode ray research, that this conductivity could occur if molecules were broken up.  But what was breaking up the air molecules and breaking them up at all levels of the atmosphere?

It is natural in scientific research to try to explain a new observation using established knowledge.  Well what sort of particles or rays were known? There was the proton, but other experiments had shown that the protons interact readily with air through what we now call the strong force; hence they would not be able to penetrate below the top levels of the atmosphere.  What about high-energy light rays, the x-rays already discovered at the end of the nineteenth century \cite{mould}.  Millikan, who had won the Nobel prize for his measurements of the electron charge, liked this hypothesis.  He pushed his hypothesis without mercy, using his power as a dominant American physicist.  But Millikan was wrong. Experiments showed that the particle or ray that made the air conductive could get through thick pieces of lead, pieces that were known to stop x-rays.  {\it Here is a lovely illustration of another uncertainty in the practice of science: great researchers can be wrong.}

By the early 1930's it was clear that mysterious particles had the ability to penetrate long distances in air and to pass through thick pieces of lead.  Since scientists name effects even when not understood, the phenomenon was called the {\it penetrating component in cosmic rays.  In the practice of science naming a phenomenon does not mean that the phenomenon is understood.}  The famous Oppenheimer even composed a theory explaining that high-energy electrons could penetrate lots of material even though it was well known that it is difficult for low-energy electrons to penetrate material.  {\it In the practice of science the very human desire to explain can lead to premature theories and wrong theories.}  Yes, Oppenheimer was wrong too.

Finally in 1937 three sets of experiments \cite{brown} reported that the penetrating component could be explained by the existence of a particle more massive than an electron but not as massive as a proton, a new particle eventually called the muon! It was almost another ten years, however, before the full nature of the muon was determined.  A complicated story had to be unraveled.  Protons and nuclei hitting the upper levels of the atmosphere produce other particles, mostly pions, through the strong force, Figure~\ref{fig:muoncos}, and the pions in turn decay into muons.  The muon does not have the strong force and so interacts very little in the air or in other matter.  Indeed it only interacts enough to make the air conducting. 

\begin{figure}[t!]
\centerline{
\epsfig{figure=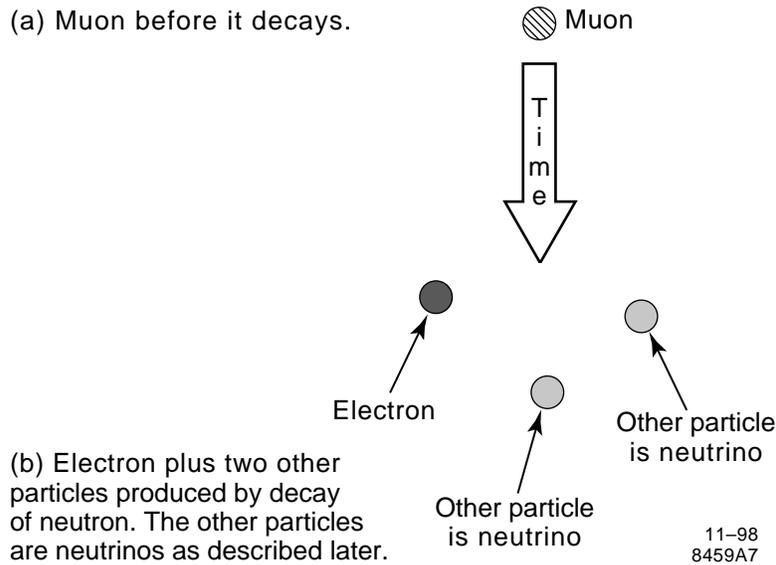}}
\caption{
The decay of a muon. (a) The muon before it decays. (b) The particles produced
by the decay.
}
\label{fig:muondec}
\end{figure}

So both the muon and the electron lack the strong force.  There is another similarity: the muon's electric charge is exactly the same as the electron's charge. Hence the muon acts electrically just like an electron.  Well why don't we have muonics, muonic motors, muonic computers?  The short answer is that the muon is unstable, it decays in about $10^{-6}$ seconds, a millionth of a second; a longer answer comes later.  The muon decays because its mass is about 200 times the mass of the electron, and the decay process, Figure~\ref{fig:muondec}, is
$$
\rm
muon \rightarrow electron + another \  particle + another \  particle.
$$
What are these other particles that occur in the decay of the muon and also occurred in the decay of the neutron?  This was the question that led to the discovery of the elementary particle, the neutrino.  

\subsection*{Theory in science: the neutrino from theory to first discovery.}

The motivation that led to the discovery of the neutrino \cite{sutten,solomey} was very different from the motivation that led to the discovery of the electron and the muon.  There was not the need to explain a general phenomenon such as cathode rays or cosmic rays; the need was to understand the experimental details of the decay process of the neutron and other nuclear decays.  Experiments had shown that the energy balance in these decays required the production of ``other particles'' but the experiments could not detect the ``other particles.''  The great theoretical physicist Pauli proposed in the early 1930's that the other particles, eventually called neutrinos, were particles with no electric charge, with no strong force, and with very small or no mass.  But such particles had never been detected.  

In the 1950's Reines and Cowan set out to see if the neutrino really existed \cite{sutten,solomey}.
They explained that their experiment was designed ``...to show that the neutrino has an independent existence, i.e. that it can be detected away from the site of its creation...''  Therefore the motivation in the narrow sense was to test an hypothesis.  In a broader sense the motivation was to see if a particle with the strange properties of the proposed neutrino could exist.  {\it This kind of motivation is very different from the phenomenon-driven motivations which led to the electron and muon discoveries.}

\begin{figure}[t!]
\centerline{
\epsfig{figure=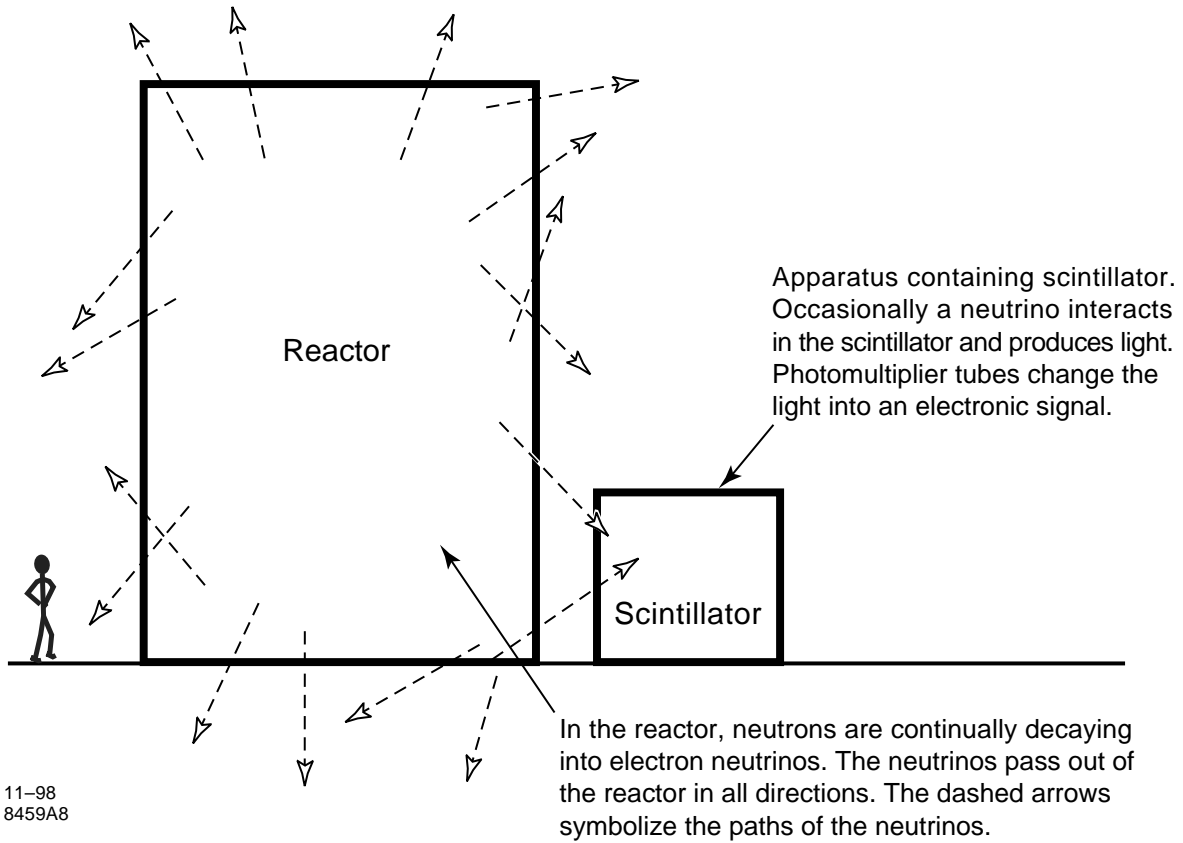}}
\caption{
The experiment of Reines and Cowan that demonstrated the existence of neutrinos.
}
\label{fig:rc}
\end{figure}

The task undertaken by Reines and Cowan was to verify the neutron decay hypothesis 
$$
\rm
neutron \rightarrow proton + electron + neutrino
$$
by showing that the neutrino existed.  But how to do this?  If the neutrino had no electric charge and no strong force it would only occasionally interact in matter through the aptly named weak force.  Their answer had two parts.  First they used a nuclear reactor in which there is a tremendous rate of neutron decay to produce an intense outflow of neutrinos, Figure~\ref{fig:rc}.  I should write a hypothetical intense outflow of neutrinos because the existence of the neutrino had not been proven.  The second part of their answer was to use a large amount of matter in the form of liquid scintillator to detect the occasional neutrino interaction in the scintillator.  A scintillator is a type of material that emits visible light when particles interact with the material.  

The experiment worked and the adjective hypothetical was removed, an accomplishment for which Reines received the 1995 Nobel Prize in Physics.  This is a classic example of scientific discovery: puzzling experimental results leading to a bold new hypothesis, and then confirmation of the new hypothesis by a new and different experiment.  But this simple sequence ignores the crucial use by Reines and Cowan of new experimental technology, the nuclear reactor and the large liquid scintillator detector.  {\it Scientific progress often depends upon the invention of new experimental technology to verify new hypotheses.}

\subsection*{Scientific laws: mass and electric charge in subatomic physics.}

In the biological, chemical and mechanical phenomena of our everyday lives the total mass never changes.  If you break a brick in two, the sum of the masses of the two pieces is equal to the mass of the original brick.  In the chemical reaction
$$
\rm
sodium \  atom + chlorine \  atom \rightarrow sodium \  chloride \  molecule
$$
the mass of the sodium chloride molecule is equal to the sum of the masses of the sodium atom and the chlorine atom.  But in the world of subatomic particles one can destroy mass, changing mass into energy.  Or one can create mass by changing energy into mass.  

An example of destroying mass is muon decay (the ``other particles'' are neutrinos): 
$$
\rm
muon \rightarrow electron + neutrino + neutrino.
$$
The sum of the masses of the electron and the two neutrinos is less than the mass of the muon.  Some of the mass of the muon has been destroyed; it has been changed into energy.  This process might be written
$$
\rm
muon \rightarrow electron + neutrino + neutrino + energy.
$$

\begin{figure}[t!]
\centerline{
\epsfig{figure=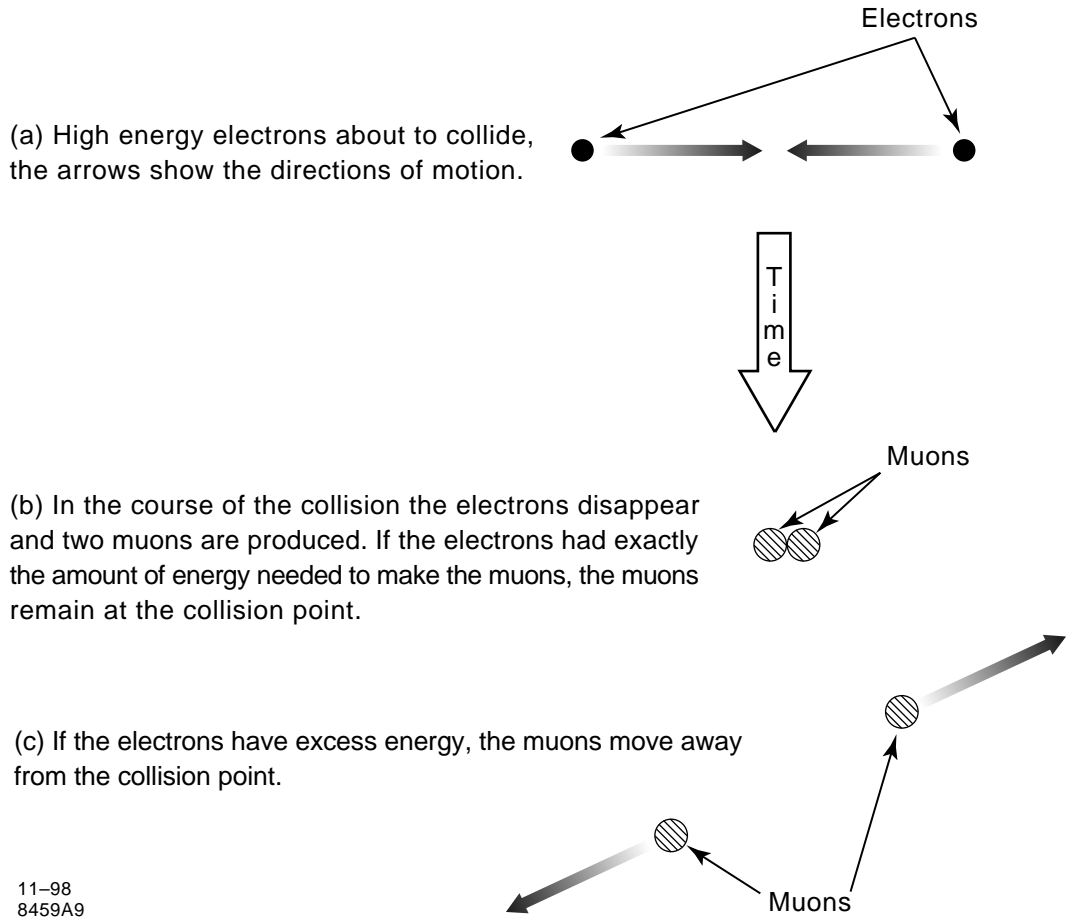}}
\caption{
The collision of two electrons leading to the process electron + electron $\rightarrow$
muon + muon. (a) Electrons about to collide. (b) Two produced muons if there is
just enough energy to make them. (c) Two produced muons moving away from the
collision point if the electrons had excess energy.
}
\label{fig:collision}
\end{figure}

The inverse process, changing energy into mass became important to experimental subatomic physics in the 1930's with the inventions and improvements of particle accelerators, a subject discussed in the next section.  An example of changing energy into mass is the reaction, Figure~\ref{fig:collision},
$$
\rm
electron + electron \rightarrow muon + muon.
$$
The mass of a muon is about 200 times the mass of an electron. Therefore in this process the electrons have to possess large amounts of energy, this energy being changed into the masses of the muons.

I want to be a little more specific about this elementary particle reaction, changing a pair of electrons into a pair of muons.  Since electrons and muons have electric charge that can be positive or negative, it is usual to specify the sign of the charge.  For example:
$$
\rm
negative \  electron + positive \  electron \rightarrow negative \  muon + positive \  muon.
$$
Negative and positive units of electric charge have the same size for all known subatomic particles, therefore on the left side of this process the negative charge exactly cancels the positive charge and the total charge going into the reaction is zero.  The products of the reaction on the right side also have total charge zero.  

But to the best of our experimental knowledge the reaction
$$
\rm
negative \  electron + positive \  electron \rightarrow positive \  muon + positive \  muon
$$
where the total amount of charge changes, {\it never occurs}.  The rule is that the total amount of charge going into a reaction must be the same as the total amount of charge coming out of the reaction.  This is called the law of the conservation of electric charge.  {\it Again beware of terminology in the practice of science. No one understands why charge cannot be created or destroyed.  Experimental rules are often called laws whether or not we understand the reason for the rule.}

On the other hand we do understand why mass can be created or destroyed in a reaction.  Mass is just another form of energy; with the right apparatus and reaction, chemical energy can be changed into mechanical energy and mechanical energy can be changed into mass energy, or one can carry out any other combination of energy changes.

\subsection*{Accelerators and high-energy physics.}

\begin{figure}[t!]
\centerline{
\epsfig{figure=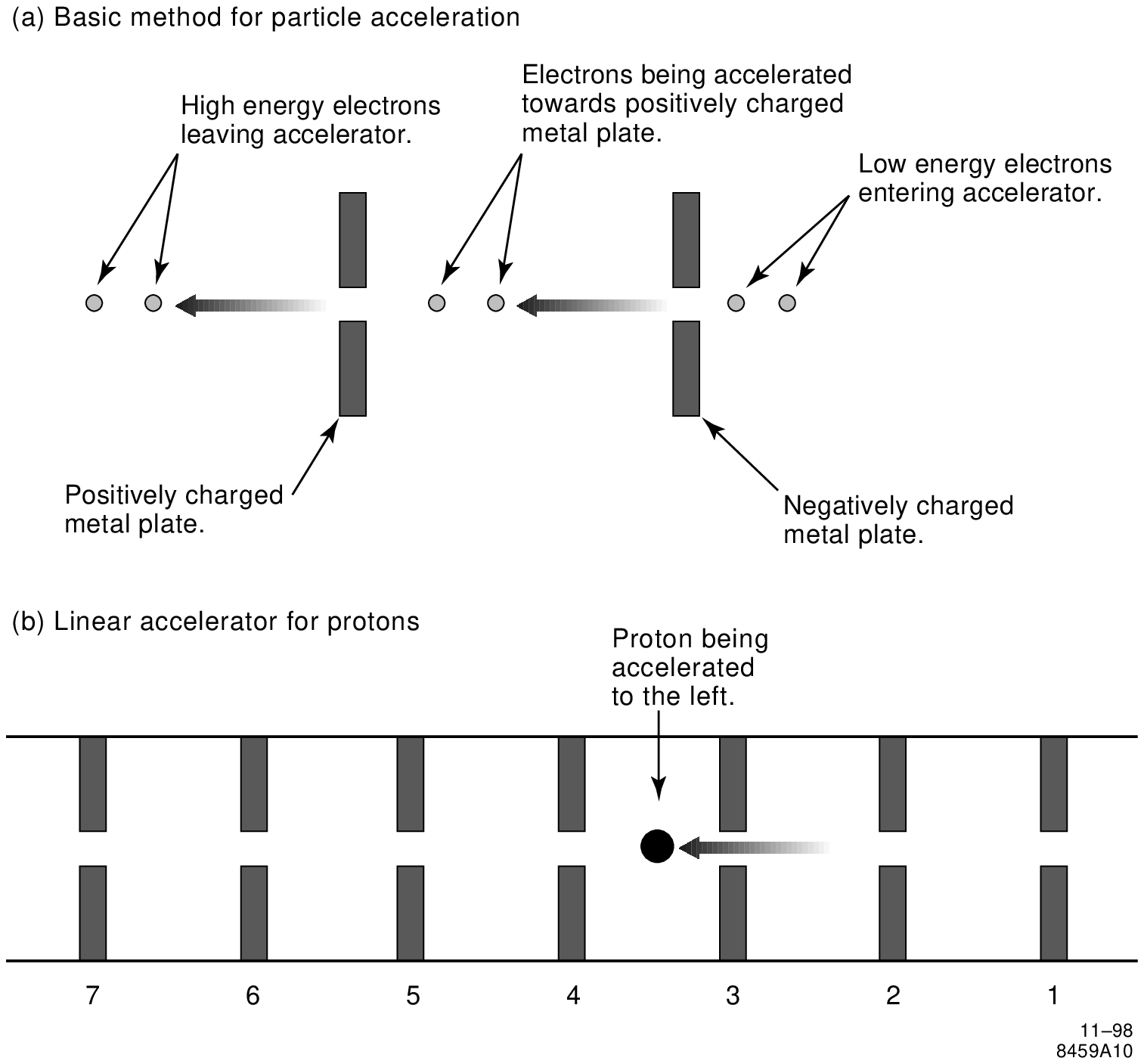}}
\caption{
(a) The basic method for particle acceleration showing how an electron gains
velocity as it is attracted by the positively charged plate and repelled by the
negatively charged plate. (b) The basic operation of a linear accelerator, here
shown for protons.  As the proton moves to the left, the plates 2, 3, 4,... in
succession are made negative, pulling the proton along and giving it more and
more velocity.
}
\label{fig:acceleration}
\end{figure}

The energy of a particle depends upon its velocity; the greater the velocity the greater the energy.  When we increase the velocity of an automobile or a particle we say we are accelerating the automobile or the particle; and so the devices that increase the velocities of particles are called accelerators \cite{hoddeson}.  The basic idea used in accelerators is simple: a charged particle will accelerate in an electric field because of the electric force on the particle, Figure~\ref{fig:acceleration}.  But accelerator technology is complicated; I show two simple, schematic examples in Figure~\ref{fig:acceleration}. 

Most present research in elementary particle physics uses high-energy particles from accelerators, and so elementary particle physics is also called high-energy physics. And indeed it can be very high energy.  For example a proton can be given so much energy that in the collision with another proton dozens of  pions can be produced:
$$
\rm
proton + proton \rightarrow proton + proton + dozens \  of \  pions.
$$

\subsection*{Two kinds of neutrinos and the lepton family.}

A major discovery using a high-energy accelerator was the experimental proof in the middle 1960's that there are two kinds of neutrinos \cite{sutten,solomey}.  One associated with the electron called an electron neutrino, \pagebreak the other associated with the muon, called the muon neutrino. Indeed in the decay of a muon these two kinds of neutrinos appear, for example:
$$
\rm
negative \ muon \rightarrow negative \ electron + muon \ neutrino + electron \ neutrino.
$$
I have put in the electric charges in these muon decays to emphasize again that total charge does not change in a reaction.  Of course the electric charge of neutrinos is zero.  I hope elementary particle physics cognoscente will forgive me for not drawing a distinction here, or in the previous sections, between a neutrino and an antineutrino.

\begin{figure}[p!]
\centerline{
\epsfig{figure=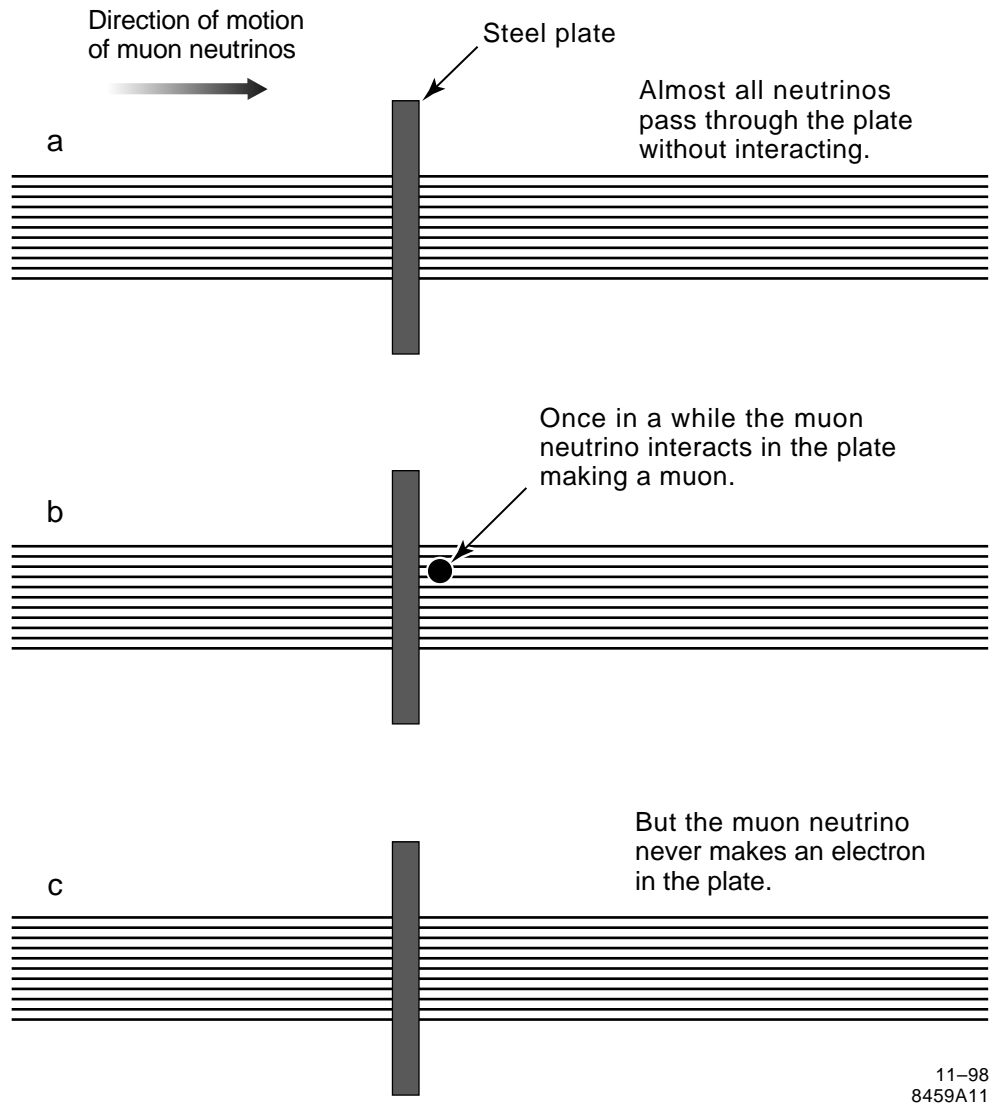}}
\caption{
The experiment of Lederman, Schwartz, and Steinberger that demonstrated the
existence of two kinds of neutrinos.  Most of the muon neutrinos do not interact in
 the steel plate as in (a), occasionally a neutrino interacts to make a muon as
in (b), but the muon neutrino never makes an electron as in (c).
}
\label{fig:lss}
\end{figure}

In an experiment for which Lederman, Schwartz, and Steinberger received the Nobel Prize, a high-energy neutrino beam was produced indirectly by a high-energy proton accelerator.  When these neutrinos interacted with matter only muons were produced, not electrons, Figure~\ref{fig:lss}.  Hence these neutrinos were muon associated.  As I have explained neutrinos rarely interact and so it was not an easy experiment.  Other later experiments have shown that one can use an accelerator to make a different kind of neutrino, neutrinos that produce only electrons when they interact with matter.

Muon neutrinos associated with muons, electron neutrinos associated with electrons, does this sound like a tautology?  It is not a tautology but the terminology promises more than we actually know.  We do know that there are these two different kinds of neutrinos, one associated with the electron, one associated with the muon.  But we do not understand the mechanism of that association. For example, is there something inside the electron that is also inside the electron neutrino?  Allowing some anthropomorphism, how does a neutrino, a particle of perhaps no size, know that it is associated with an electron, another particle of perhaps no size?  {\it The lesson here about the practice of science is again that terminology may appear to have deeper meaning than is warranted.}

By the middle 1960's the electron, muon, and the two neutrinos were thought of as forming a family, the lepton family, Table~\ref{table:chargedlep}.  The identification of the lepton family was based on two considerations.  First, these four particles have nothing to do with the strong force. Second, they had very small masses compared to most other subatomic particles, less mass than the proton, less mass than the neutron, and less mass than other subatomic particles such as the pion.  The Greek word {\it leptos} means small or fine, the electron, muon and two neutrinos were thought to be the smallest mass subatomic particles.  {\it Again not a profound terminology.}
\begin{table}[t]
\begin{center}
\caption{
\label{table:chargedlep}
The properties of the charged leptons and their associated neutrinos.  The electric charge of the charged leptons is $1.6 \times 10^{-19}$ coulombs.  The neutrinos, have zero electric charge.  The masses of the leptons are given in terms of the electron mass, for example the muon has 207 times the mass of the electron.  At present we are sure of only upper limits on the neutrino masses, but we may be on the verge of knowing more about neutrino masses, as discussed at the end of the paper.  The actual mass of the electron is about $10^{-27}$ grams.  For additional information on lepton properties see \protect\cite{kane}.
}
\vspace{3ex}
\small
\begin{tabular}{|p{.34\textwidth}||p{.2\textwidth}|p{.18\textwidth}|p{.16\textwidth}|}
\hline
Charged lepton name&Electron&Muon&Tau\\
\hline
\hline
Years from first research to final discovery&1860 to 1895&1925 to 1945&1970 to 1978\\
\hline
Charged lepton symbol&$e$&$\mu$&$\tau$\\
\hline
Charged lepton mass in terms of electron mass&1&207&3480\\
\hline
Charged lepton lifetime in seconds&stable&$2.2 \times 10^{-6}$&$2.9 \times 10^{-13}$\\
\hline
Associated neutrino name&electron neutrino&muon neutrino&tau neutrino\\
\hline
Associated neutrino symbol&$\nu_e$&$\nu_\mu$&$\nu_\tau$\\
\hline
Upper limit on mass of associated neutrino in terms of electron mass&less than 1/50,000 of electron mass&less than 1/3 of electron mass&less than 40 times electron mass\\
\hline
Upper limit on mass of associated neutrino in terms of its charged lepton mass&less than 1/50,000 of electron mass&less than 1/600 of muon mass&less than 1/90 of tau mass\\
\hline
\end{tabular}
\end{center}
\end{table}
\normalsize

There is one other particle that also has a very small mass, probably exactly zero mass.  Sometimes light interacts with matter as a particle, not as a light wave, and this particle is called a photon.  I mentioned the photon at the beginning of the paper as belonging to the third family of elementary particles, particles that carry the basic forces.  In this paper, I cannot give an explanation useful to the reader of the difference between the lepton family and this third family.  All I can say is that the photon behaves very differently from the electron, muon, and neutrinos; to include it in the lepton family would destroy the meaning of the lepton classification.  

\subsection*{Applications of basic research: electrons, muons, and neutrinos.}

The electric telegraph was developed and put into use in the in the first half of the nineteenth century \cite{dunsheath} long before the discovery of the electron, even though its operation depends on the properties of electrons in metals.  The same is true of the electric motor, electric generator, telephone, and electric lights; all were invented and used before the discovery of the electron, although all depend on the properties of the electron.  Even the early days of wireless technology were not based on electron physics \cite{lewis}. {\it The invention and use of new technology need not depend on basic research; in fact for most of history it has not depended on basic research.}

With the discovery of the electron and the understanding of its behavior in solids, gases, and vacuums, however, many more inventions were possible.  Radio technology, until the commercial use of the transistor in the 1950's, depended upon the vacuum tube in which the electric current is carried through the vacuum by electrons going from the cathode to the anode, the old cathode ray idea.  Of course the transistor, the integrated circuit, and the computer all depend upon a thorough understanding of the behavior of electrons in semiconductor metals such as silicon \cite{riordan}.  {\it And so we have become very use to the idea that basic research can lead to new technology: the wonder of the computer, the horror of nuclear weapons.}

I wrote earlier that we cannot use muons in electron technology because the muons are unstable.  In addition muons behave very differently in matter from electrons.  {\it New discoveries sometimes can be used to duplicate or improve old technology, but more often new discoveries lead to new technology or no technology.}

There is a possibility that muons might someday be used in a new energy technology.  The sun produces energy by the fusion of nuclei.  As a simplified example, if a proton fuses with a deuteron, a combined nucleus plus energy is produced, Figure~\ref{fig:fusion}:
\begin{figure}[p!]
\centerline{
\epsfig{figure=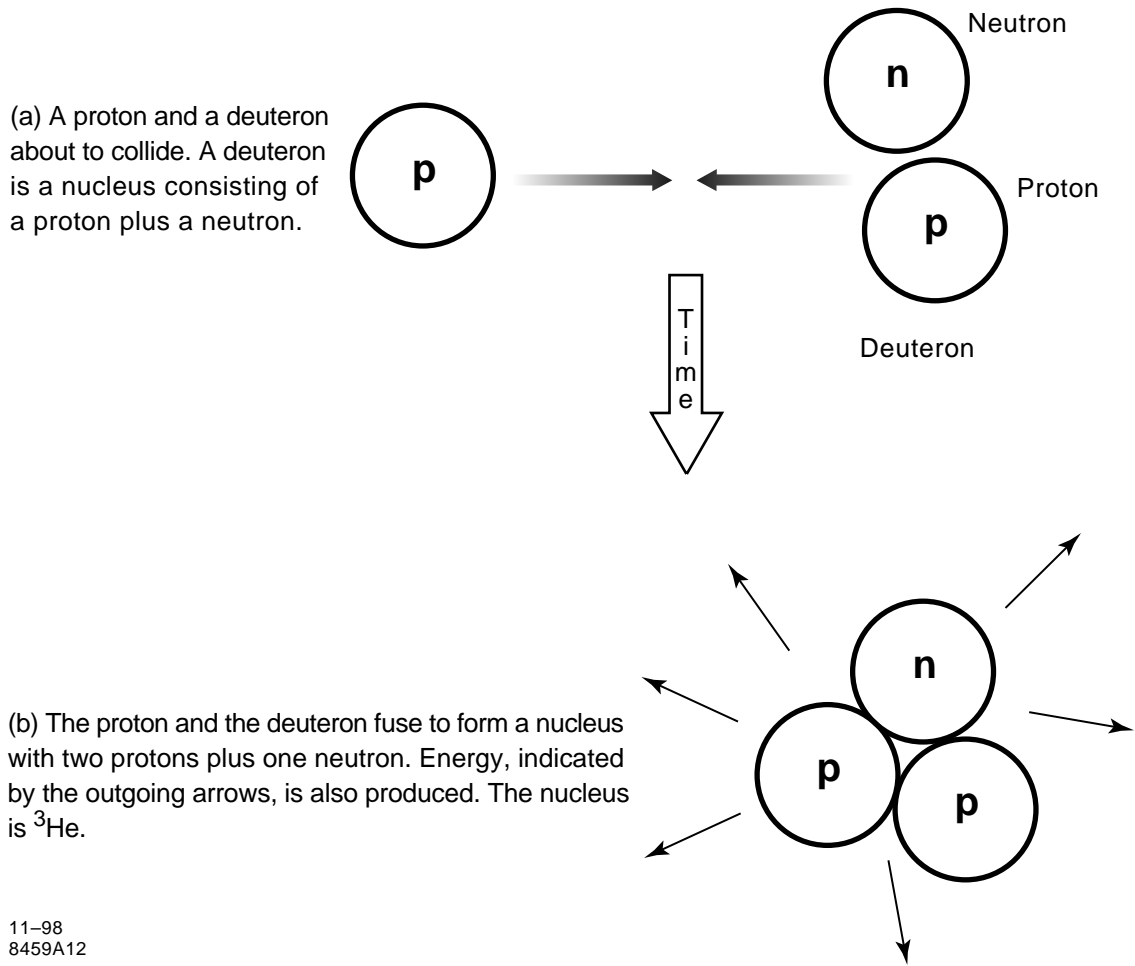}}
\caption{
Simplified example of energy production by high-temperature nuclear fusion.
 In (a) a proton and a deuteron are about to collide because of the velocities
they have acquired via high temperature. In (b) after the collision, the proton
and deuteron fuse into a $^3$He nucleus and also produce energy.  A deuteron is a
nucleus consisting of a proton and a neutron; it is the nucleus of heavy hydrogen.
This is a simplified example of the fusion reactions that produce energy in
our sun, in stars, and in the fusion (usually called hydrogen) bomb.
}
\label{fig:fusion}
\end{figure}
$$
\rm
proton + deuteron \rightarrow combined \  nucleus + energy.
$$
A proton is the nucleus of the hydrogen atom and a deuteron is the nucleus of the hydrogen atom found in heavy water.  Both exist naturally, but the problem is getting the proton and the deuteron to collide with sufficient force to fuse.  The high temperature of the sun and stars produces the required forceful collisions.  By the way, this high temperature requirement is also the reason for the lack of success in producing energy on Earth through controlled fusion.  

\begin{figure}[p!]
\centerline{
\epsfig{figure=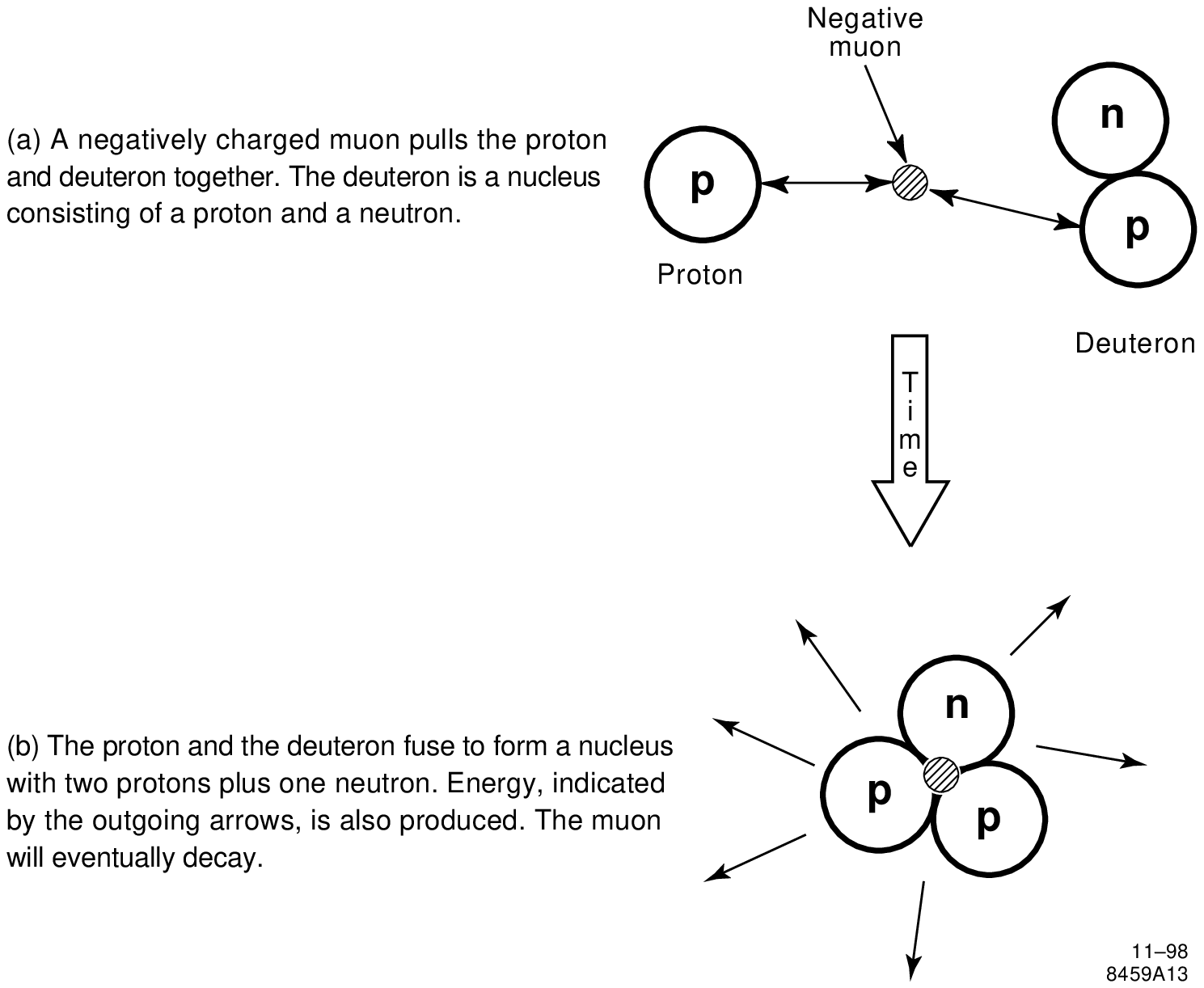}}
\caption{
Simplified example of energy production by muon-induced nuclear fusion.  In
(a) a proton and a deuteron are being pulled together by a negatively charged
muon. In (b) after the collision, the proton and deuteron fuse into a $^3$He nucleus
and also produce energy.  The muon eventually decays.  This process has been
demonstrated experimentally, but it is not at present commercially feasible
because too much energy is required to produce the muons.
}
\label{fig:fusionmuon}
\end{figure}

But negative muons can produce proton-deuteron fusion at room temperature.  This was demonstrated and understood decades ago.  The negative charge of the muon pulls together the positively charged proton and positively charged deuteron; no extra temperature is needed, Figure~\ref{fig:fusionmuon}.  Where then are the muon fusion reactors?  The problem is that muons decay, and so new muons have to be continually created using an accelerator.  With existing accelerator technology the energy required to produce the muons is greater than the energy from the fusion.  No one has yet designed a sufficiently efficient accelerator, but it may not be impossible to do so.  {\it Sometimes new discoveries can in principle lead to new technology, but economics or unfortunate technological problems may prevent practical use.}

What about the neutrinos? Do they offer any practical uses?  There have been suggestions that muon neutrino beams could be used for geological research and prospecting deep in the Earth.  The rate of interaction of the neutrinos would be proportional to the density of matter; the interaction rate being measured by the muons so produced.  A grand but futuristic engineering project.  

\subsection*{Puzzles in science: the electron-muon problem.}

I now come to my research in lepton physics.  In 1963 I joined the Stanford Linear Accelerator Center, SLAC, to do research in high-energy physics.  I had been working with particles that interacted through the strong force, a broad and popular, but complicated, area.  I wanted to work in a simpler area and so my thoughts turned to the known leptons before 1970: the electron, the muon, and the two neutrinos, Table~\ref{table:chargedlep}.  I was particularly intrigued by the connection between the electron and muon.  With respect to the electromagnetic force, and the absence of the strong force, the muon behaves simply as a heavier electron, 207 times heavier.  But why is it 207 times heavier?  

Another puzzle was understanding the muon decay 
$$
\rm
negative \ muon \rightarrow negative \ electron + muon \ neutrino + electron \ neutrino.
$$
Why doesn't the muon decay to the electron through the simpler process
$$
\rm
negative \ muon \rightarrow negative \ electron + photon \ ?
$$
The photon has very small or zero mass just like the neutrinos, and the photon has zero electric charge so that the charge is the same on both sides of the decay reaction.  By the middle 1960's these questions and puzzles were called the {\it electron-muon problem}.

I thought that SLAC would be an excellent place to work on the electron-muon problem.  A high-energy electron accelerator was being built at SLAC, intense beams of high-energy electrons were available, and SLAC researchers were starting experiments on the collision of electrons with protons and nuclei.  It was also easy to use the electrons to make beams of high-energy muons.  I decided to start high-energy experiments on the collision of muons with protons and nuclei.  

\subsection*{Obsession in science.}

I was obsessed with a simple idea.  Since the muon is much heavier than the electron, I speculated that the muon somehow had some of the heavier proton's nature.  Therefore I thought that the collisions of muons with protons would be different from the collisions of electrons with protons.  Don't try to follow this idea in detail because after five years of experiments with muons I realized in the early 1970's that I should give up this idea.  

My colleagues and I found that there would always be errors of 10 or 15\% in comparisons of our muon-proton collision measurements with the electron-proton collision measurements of the other SLAC researchers.  We knew of no way to improve the precision of our experiments.  And so even though obsessed with the electron-muon problem I gave up.  {\it It is important in the practice of science to know when to be obsessed and when to give up the obsession; it is important to learn the art of scientific obsession.  A scientist needs obsession to keep going through the ups and downs of research, but obsession can also lead to years or decades of pointless research.} There should be some comfort in the thought that if an idea is good, scientists will return to it in future years with better experiments and better theoretical understanding.  

To close this part of the story, since the early 1970's there have been many better experiments on muon-proton collisions, experiments carried out for other reasons.  These experiments have shown that nothing can be learned about the electron-muon problem from such experiments.  Thus it was lucky that I gave up the obsession.  {\it Of course sometimes one gives up a scientific obsession, only to find that years later others have made it pay off.}

My attack on the electron-muon problem being thwarted by nature, I turned to another idea that had been in my mind since the early 1960's.  Perhaps there was another undiscovered and more massive charged lepton, a heavy relative of the electron and the muon.  As my attack on the electron-muon problem began to go badly, I became more and more optimistic about finding a new charged lepton.  As Voltaire wrote in {\it Candide}, ``Optimism, said Candide, is a mania for maintaining that all is well when things are going badly.''

\subsection*{Simple ideas in science: my search for a new lepton.}

I became obsessed with another simple idea.  I thought that the electron and the muon might be the smallest mass members of a much larger family of leptons.  I drew for myself the following chart:

\small
\begin{center}
\begin{tabular}{p{.249\textwidth}|p{.37\textwidth}|p{.31\textwidth}}
\hline
We know that there is:&an electron &with its associated neutrino.\\
We know that there is:&a muon heavier than electron &with its associated neutrino.\\
Perhaps there is:&a lepton\#3 heavier than muon &with its associated neutrino?\\
Perhaps there is:&a lepton\#4 heavier than lepton\#3 &with its associated neutrino?\\
And so forth&&\\
\hline
\end{tabular}
\end{center}
\normalsize

\noindent Thus I dreamed that there were a large number of more and more massive charged leptons: electron, muon, lepton\#3, lepton\#4, lepton\#5 and so forth; and that each of these charged leptons had an associated neutrino.  Why such a large number?  Because I didn't see why there should be any upper limit to the mass of a lepton.

I had a hidden motivation in this search for new leptons.  No one had solved the electron-muon problem; there were not enough clues.  But if an additional charged lepton were to be discovered, then we would have the electron-muon-lepton\#3 problem.  We would surely have more clues.  {\it A basic principle in the practice of science: if you can't solve a problem get more data.}

I decided the best way to look for these additional charged leptons was to copy the old reaction
$$
\rm
negative \ electron + positive \ electron \rightarrow negative \ muon + positive \ muon.
$$
We could look for new lepton\#3 by using the reaction
$$
\rm
negative \ electron + positive \ electron \rightarrow negative \ lepton\#3 + positive \ lepton\#3.
$$
Of course we would be delighted to find any new lepton, lepton\#3 or lepton\#4 or lepton\#5. 

It was a very optimistic proposal: these hypothetical leptons had to exist and our experimental equipment had to work well enough for us to find them.  An obvious worry was that the new leptons might exist, but we might not have enough energy to produce the required masses.  

I followed a rule of mine for starting new science ventures.  {\it If you get a new idea for an experiment, don't spend forever trying to understand every detail of how you will carry it out, just start.  You will learn as you proceed with the experiment.} There is a problem in this rule.  {\it Usually one gets five or ten bad or fruitless ideas for every good idea. This means that you will spend much time on bad ideas.  Unfortunately in the practice of experimental physics, it usually takes time to identify the good idea.}

\subsection*{The discovery of the tau lepton.}

Most people in the high-energy physics community of the early 1970's didn't know or didn't care about this search for new leptons.  Of those who did know about the search, most were skeptical, even among my colleagues.  {\it In the practice of science there is often a choice between working in a popular area that most colleagues feel is fruitful or working in an unpopular area that most colleagues feel is a waste of time and research money.  Most of the time the popular field is the fruitful field, but a discovery in an unpopular area brings more satisfaction and more fame.  In the end it is a question of one's personality.}

\begin{figure}[t!]
\centerline{
\epsfig{figure=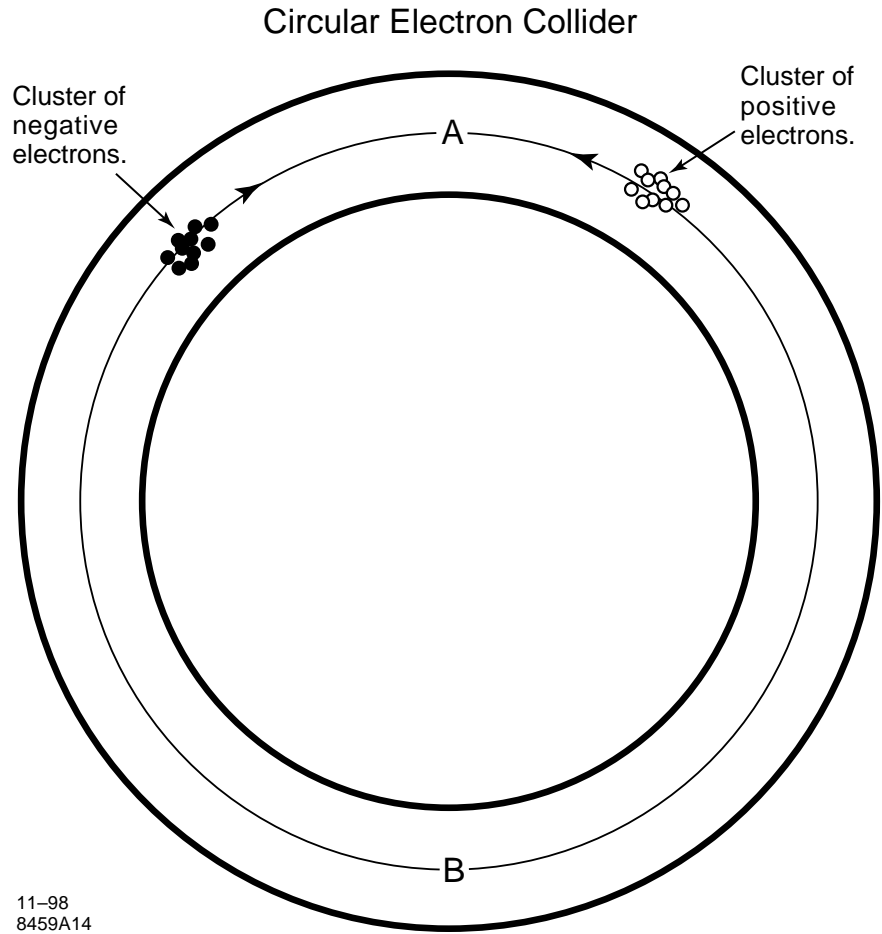}}
\caption{
Schematic diagram of a circular electron collider.  A cluster of high-energy
 negative electrons and a cluster of high-energy positive electrons move along a
 circular path in opposite directions as shown by the arrows.  The clusters meet
 at points A and B.  When the clusters meet most of the electrons pass by each
other, but sometimes a negative electron collides with a positive electron.  Then
 reactions occur such as electron + electron $\rightarrow$ muon + muon and electron + electron $\rightarrow$ tau + tau.
}
\label{fig:collider}
\end{figure}

About 1970 we were completing at SLAC the construction of an electron collider called SPEAR.  Electron colliders, Figure~\ref{fig:collider}, were a new technology in high-energy physics; their development began in the 1960's.  Electron colliders provide the means to collide negative electrons with positive electrons at high energy with great intensity.  The work I am about to describe could not have been done without the new technology of electron colliders.  {\it As I have already written, in the practice of science new technology is often crucial.}

In 1973 the SPEAR electron collider began operation and my colleagues and I began to look for a new lepton.  By 1975 we began to find evidence for the existence of a third lepton, a lepton much more massive than the muon, in fact about 17 times more massive!  My close colleagues and I were excited, delighted, and overjoyed.  We were detecting the reaction
$$
\rm
negative \ electron + positive \ electron \rightarrow negative \ lepton\#3 + positive \ lepton\#3.
$$
We knew it was lepton\#3 because our experiments and other experiments had shown there was no other lepton with a mass greater than the muon but less than that of lepton\#3.

But the larger elementary particle physics community remained skeptical, doubting and criticizing our research.  The problem was that while we continued to find more and more evidence for the existence of a new lepton, other experimenters could not verify our results using somewhat similar experimental methods.  Furthermore a few of these experimenters were not eager to find verification.  {\it A somewhat unpleasant consequence of scientific competition is that once a discovery is claimed, others can get more credit for disproving the claim than for verifying the claim.}  It was a tough few years for me; our evidence kept growing but there was no outside verification.  {\it Certainty in science comes from verification of one's findings by others; this is fundamental scientific practice. It allows the eventual overcoming of uncertainties in scientific practice.}

Finally in late 1977 other experimenters began to find our new lepton.  We gave lepton\#3 the Greek name tau, because tau, written as $\tau$, is the first letter of the Greek word for third.  And so the reaction becomes
$$
\rm
negative \ electron + positive \ electron \rightarrow negative \ tau + positive \ tau.
$$

\subsection*{The road from difficult research to easy research: the tau lepton.}

Our discovery was based upon studying about one hundred examples of this reaction and the properties of the tau leptons so produced.  Today I work with an experiment at Cornell University where millions of these reactions have been detected and millions of tau leptons have been studied.  An improved electron collider at Cornell, and new electron colliders at my laboratory SLAC and at the KEK laboratory in Japan, will enable physicists to study ten million examples per year of
$$
\rm
negative \ electron + positive \ electron \rightarrow negative \ tau + positive \ tau.
$$
{\it Usually improvements in technology allow obscure and difficult scientific studies to become easier and easier.  This adds to the certainty of scientific results.  If studies of a phenomenon never get easier, if the technology for carrying out the studies never improves, then it is most probably not science that is being practiced.}  Thus for psychic phenomena: it is no easier to verify the reality of telepathy than it was five hundred years ago.

A few notes on the properties of the tau.  Its mass is 3480 times the mass of the electron, Table~\ref{table:chargedlep}.  It has the same size electric charge as the electron, and like the electron and muon, it has nothing to do with the strong force. It is indeed a lepton. With the discovery of the tau, however, the name lepton has lost its original meaning.  The tau is not a light particle it is a heavy particle having about twice the mass of the proton.  There is a neutrino associated with the tau called, of course, the tau neutrino.  

The tau, like the muon, is unstable. It decays in an average time of roughly $10^{-13}$ seconds. There are many ways in which the tau decays, but two of these ways beautifully demonstrate connections between the electron, the muon, and the tau. Recall that the muon decays through the process
$$
\rm
negative \ muon \rightarrow negative \ electron + electron \ neutrino + muon \ neutrino.
$$
Two of the ways the tau decays are
$$
\rm 
negative \ tau \rightarrow negative \ electron + electron \ neutrino + tau \ neutrino
$$
$$
\rm
negative \ tau \rightarrow negative \ muon + muon \ neutrino + tau \ neutrino.
$$

\subsection*{Nature can be cruel: the electron-muon-tau problem.}

I had dreamed that once a new lepton was found, the properties of the new lepton would provide new clues to the inner nature of leptons, indirectly solving the old electron-muon problem.  Now in 1998 we know a tremendous amount about the properties of the tau lepton.  There have been hundreds of physics Ph.D. theses on the properties of the tau, more than a thousand experimental and theoretical papers on the tau, and every two years we hold an international conference devoted solely to the tau.  But there are no new clues to the inner nature of the leptons.  If you assume that the tau behaves exactly like a heavier version of the muon, and if you use knowledge acquired in other parts of subatomic physics, you can predict quantitatively the behavior of the tau.

From one point of view this is wonderful, it shows that we are developing certain and consistent understanding of the behavior of elementary particles.  But from the point of view of those who want to push deeper into the world of elementary particles, who want to push below the bottom of Figure~\ref{fig:hierarchy}, this is disheartening.  {\it Nature can be cruel.}

\subsection*{The uncertainty of research directions: are there more leptons?}

If you look back a few pages you will see that I dreamed not only of lepton\#3, but also lepton\#4 and lepton\#5 and so on.  Since the discovery of the tau there have been many, many searches for additional leptons.  Yet no more have been found.  I am as surprised as anyone.  The powerful method we used to discover the tau 
$$
\rm
negative \ electron + positive \ electron \rightarrow negative \ tau + positive \ tau
$$
has been used at ever-increasing energies to search for the next charged lepton
$$
\rm
negative \ electron + positive \ electron \rightarrow negative \ lepton\#4 + positive \ lepton\#4.
$$
As I write this paper these searches have been carried out at the CERN European laboratory up to energies more than 50 times greater than the energy at which we discovered the tau.  No new charged lepton has been found.  This means that if lepton\#4 exists, its mass is larger than about 50 times the mass of the tau, about 180,000 times the mass of the electron.  In addition by a related method, experimenters have searched for heavier neutrinos; nothing has been found beyond the three known neutrinos, those associated with the electron, muon, and tau.  The searches at CERN are not completed; they will extend about 10\% higher in energy.  {\it Sometimes in the practice of science, changing one parameter in an experiment can lead to a discovery; it can be a change in energy or in precision or in the amount of data.}

About ten years from now, perhaps a little later, a new kind of electron collider called a linear collider will go into operation.  This new technology accelerator is being developed at SLAC, in Japan, and in Europe.  It will produce five times more energy than existing electron colliders.

For the present there are two possibilities.  One possibility is that there are no more leptons beyond the six in Table~\ref{table:chargedlep}.  That's bad because at present we don't understand why the number of lepton types is limited and we may not get any more clues from studying the leptons themselves.  {\it But there are smart young women and men entering high-energy physics; they may bring us that understanding.}

The other possibility is that there are more leptons, and we just don't know how to find them.  Each of the known leptons was discovered using a different experimental technology.  The electron was found in the cathode ray phenomenon, the muon was found in cosmic rays, the electron neutrino was found using a reactor, the difference between the electron neutrino and the muon neutrino was discovered using a high-energy proton accelerator, and the tau was found using an electron collider.  Perhaps this was simply because the discoveries stretched over a hundred years and technologies keep changing; or perhaps leptons are so elusive that a new technology is required for each discovery.

Perhaps the next charged lepton is so massive that it is beyond the energy reach of present or near future searches using electron collider technology, even searches using the projected linear colliders.  {\it Thus there is an uncertain future for the hundred years of research on new leptons.  We may have to give up much hope of finding new leptons, or we have to find a new technology.  This is how a research direction can become uncertain even though it has been fruitful.}

\subsection*{Speculative experiments and the practice of science.}

But I have not given up; I have been speculating about other possibilities.  In fact my colleagues and I are carrying out experiments based on these speculations.  Perhaps there is a new type of massive, charged lepton that already exists in nature.  Suppose this new type lepton was stable like the electron and had been produced in the early universe, perhaps in the ``big bang.''  Then it might be present in old pieces of matter such as meteorites and ancient rocks.  For convenience I am going to call this hypothetical new lepton the lambda.
But if the lambda is massive, why should it be stable, why shouldn't it decay like the muon and the tau:
$$
\rm
negative \ lambda \rightarrow negative \ muon + muon \ neutrino + lambda \ neutrino
$$
$$
\rm
negative \ lambda \rightarrow negative \ electron + electron \ neutrino + lambda \ neutrino \ ?
$$
These decays would be prevented and the lambda made stable by two kinds of speculative changes in the usual properties of a charged lepton.  One speculation is that the lambda does not have the usual size of electric charge, but has some fractional electric charge, say 1/2 of the usual charge or 5/4 of the usual charge.  Then the decays written above would not occur because the electric charge would be different after the decay compared to the electric charge before the decay.  And as far as we know the total electric charge cannot change in a reaction.  

The other speculation is based on the observation that the muon and tau need their associated neutrinos in order to decay.  If one assumes that there is no neutrino associated with the lambda, then its decay might be prevented.

My colleagues and I at SLAC are carrying out experiments searching for massive, stable leptons.  We are not using accelerators; we are using a highly automated and modernized version of the apparatus used by Millikan ninety years ago to measure the electron's charge \cite{perlam}.  {\it If one is going to engage in a speculative experiment, there are three criteria that should be satisfied.  One should make sure that the speculation does not violate established scientific knowledge.  Carrying out the experiment should be interesting and pleasurable, that may be the only reward.  It should be easy for others to duplicate the experiment so that the verification of a speculation can be checked.}

\subsection*{Neutrino masses and a surprising return to cosmic rays.}

I am about finished with my recounting of a hundred years of lepton research and what it teaches us about the practice of science.  There is one more episode having to do with the masses of the neutrinos.  It has been very hard to measure these masses; we only know for certain the upper limits given in Table~\ref{table:chargedlep}, and there is even the possibility that neutrinos have zero mass.

For decades it has been suspected, or at least hoped, that neutrinos might change into each other, a muon neutrino change into an electron neutrino, or the converse, an electron neutrino change into a muon neutrino, or a muon neutrino change into a tau neutrino.  If such changes could occur, then a general principle of quantum mechanics predicts that the rate of change depends on the masses of the neutrinos.  In the last decade there have been many searches for this neutrino-changing phenomenon.  Experimenters have used electron neutrinos from reactors, the same neutrinos that Reines and Cowan first detected.  Experimenters have used muon neutrino beams from high-energy accelerators, the same sort of beam used by Lederman, Schwartz, and Steinberger to show that there are two kinds of neutrinos.  But all these experiments have been inconclusive at best.

Now, as the century ends, the phenomenon of neutrino-changing may have been finally detected in of all places, cosmic rays.  One effect of cosmic rays is to produce muon neutrinos, and these muon neutrinos pass through the atmosphere and into the Earth.  We know enough about cosmic rays to predict how many muon neutrinos should hit the Earth's surface per second.  A vast new underground apparatus in Japan, called Super-Kamiokanda, has been used to count the number of muon neutrinos, and there seem not to be enough of them \cite{science}.  Furthermore, it seems as though the missing muon neutrinos have changed into tau neutrinos or into some unknown neutrino, but not into electron neutrinos.  This means that the muon neutrino and perhaps the tau neutrino definitely have a non-zero mass.  But it could be a very small mass, less than 1/1,000,000 of the electron mass.  

These first results require verification from other experiments looking at cosmic rays and elucidation from experiments using reactors or accelerators.  Still the results demonstrate the surprises that can occur in science.  {\it Surprises are the best part of the practice of science, but most surprises require the experimenters to do something new and different, such as examining a new phenomenon or applying a new technology to an old phenomenon.}

\vspace{-.5ex}
\subsection*{Looking ahead.}
\vspace{-.5ex}

An up-to-date physicist in 1899 would have known about the electron and some of its properties, but would have not been able to know anything else about the rest of the world of lepton physics.  We are in the analogous state of ignorance in 1999.  Whatever the science--physics, chemistry, biology, psychology--we cannot know what we will learn in the next hundred years.  We only know that the practice of science is full of uncertainties and that the test of reality is always experiment and observation.  Darwin wrote, ``I must begin with a good body of facts and not from a principle (in which I always suspect some fallacy) and then as much deduction as you like.''

\vspace{-.5ex}
\subsection*{Appendix on very large and very small numbers.}
\label{appendix}
\vspace{-.5ex}

It is tedious to write and hard to decipher a very large number such as 100,000,000,000.  It is better to use the notation $10^{\rm N}$ where N tells us the amount of zeros in the number.  For example:

\indent\indent\indent\indent\indent\indent\indent\indent One thousand = 1,000 = 10$^3$\\
\indent\indent\indent\indent\indent\indent\indent\indent One million = 1,000,000 = 10$^6$\\
\indent\indent\indent\indent\indent\indent\indent\indent Ten million = 10,000,000 = 10$^7$.\\

A number such as $1.5 \times 10^7$ means $1.5 \times 10,000,000$. 

An analogous system is used for very small numbers such as 1/100,000. The number is written $10^{-\rm N}$, where the negative sign indicates that the number of zeros that are in the denominator.  Thus\\
\indent\indent\indent\indent\indent\indent\indent\indent\indent\indent 1/1,000 = $10^{-3}$\\
\indent\indent\indent\indent\indent\indent\indent\indent\indent\indent 1/1,000,000 = $10^{-6}$.\\

A number such as $1.5 \times 10^{-6}$ means 1.5/1,000,000.

\pagebreak

\end{document}